\documentclass[12pt]{article}
\makeindex

\usepackage{graphicx}
\usepackage{amsmath,amsthm,latexsym,amssymb,amsfonts,epsfig,MnSymbol,proof}

\newcommand{\aaa}{\mathcal{A}}
\newcommand{\ajm}{\mathbf{A}}
\newcommand{\asi}{A}

\newcommand{\capb}{\mathbf{B}}

\newcommand{\gnn}{\mathcal{N}}
\newcommand{\ppx}{X}

\newcommand{\qs}{Q}
\newcommand{\rlv}{R}

\newcommand{\eventa}{\mathcal{F}}
\newcommand{\samps}{\Omega}

\newcommand{\nrlv}{\overline{\rlv}}

\newcommand{\qss}{\bra{\qs}}

\newcommand\Sets{{\bf Sets}} 

\newcommand{\bra}[1]{\pmb{\langle}#1\,\pmb{|}}

\begin{document}

\title{Factory of realities: on the emergence of virtual spatiotemporal structures}\label{zapatrin}

\author{Rom\`an R. Zapatrin\footnote{Informatics Dept., The State Russian Museum, In\.zenernaya 4, 191186, St.Petersburg, Russia, e-mail: Roman.Zapatrin at Gmail dot com}}

\date{}

\maketitle

\begin{abstract}
The ubiquitous nature of modern Information Retrieval and Virtual World give rise to new realities. To what extent are these `realities' real? Which `physics' should be applied to quantitatively describe them?  In this essay I dwell on few examples. The first is Adaptive neural networks, which are not networks and not neural, but still provide service similar to classical ANNs in extended fashion. The second is the emergence of objects looking like Einsteinian spacetime, which describe the behavior of an Internet surfer like geodesic motion. The third is the demonstration of nonclassical and even stronger-than-quantum probabilities in Information Retrieval, their use.

Immense operable datasets provide new operationalistic environments, which become to greater and greater extent ``realities''. In this essay, I consider the overall Information Retrieval process as an objective physical process, representing it according to Melucci metaphor in terms of physical-like experiments. Various semantic environments are treated as analogs of various realities. The readers' attention is drawn to topos approach to physical theories, which provides a natural conceptual and technical framework to cope with the new emerging realities. 
\end{abstract}

\section*{Introduction}

The idea to treat Information Retrieval, or, in general, data and knowledge proceeding as physical processes has long tradition, in particular, these ideas gave rise to Quantum Computation. In the last decades a research was carried out on viewing usual (non-quantum) information deals as physical processes \cite{df82,Grib90,LPQM}. More than 100 years ago along with the growth of the precision of physical measurements Quantum Mechanics was born. Now, along with the development of World Wide Web and the emergence of immense data corpora, new conceptual challenges arise. The amount of the existing data together with the possibility to generate new data on-the-fly give rise to purely theoretical inquiries whose practical goals was to boost the performance of search engines. While the notion of ``performance'' in this context remains ambiguous, the efforts were made to build a general framework for Information Retrieval in newly emerged environment. It was observed -- in practice, by trials -- that the performance of search algorithms may be sometimes improved if one follow a ``wrong'' probabilistic model, that is, recalculate probabilities in discordance with Kolmogorovian laws. Similar situation takes place in Quantum Theory, which gave rise to Quantum Probability. Furthermore, quantum probabilistic approach was successfully applied to Information Retrieval \cite{vanRijsbergen04}. 

The development of World Wide Web, the emergence of massive accessible data sets gave rise to a kind of new realities. Like conventional physics, Information Retrieval deals with events and, as it was recently discovered, the statistical dependencies observed (that is, obtained in experiments) in Information Retrieval processes may be not only non-classical, but also demonstrate stronger-than-quantum correlations. From a broader conceptual perspective, Information Retrieval in its current state can be treated as a factory producing various realities. I intentionally reduce the picture to that extent that from the point of view of `end user', modern sophisticated physical laboratory is just a man-computer interaction. The User affects the Environment, the Environment reacts somehow, the User, being to this or that extent satisfied (this is to be quantified as well) refines its query according to some underlying principles. 

The above mentioned `underlying principles' are based on certain picture of reality which need not be classical, or even quantum. The suggested picture of operationalistically perceived Environment requires a consistent realistic framework. In recent years, a formalism based on topos theory was developed meeting this requirement. In this essay, the appropriate mathematical models based on topos theory and giving rise to event structures similar to those forming traditional spacetime, are overviewed.

In the meantime, the attempts to build a consistent and in some sense ``realistic'' theory of quantum gravity faced severe problems. The reason was that the very notion of a single, pre-defined underlying configuration space is strange for the theory, it required a formalism dealing with varying or emerging spacetime. The problems in both Information Retrieval and Quantum Gravity were of similar kind -- how to make space and time secondary, derived, rather than fundamental entities. In order to explain the observed non-classicality, the idea of non-uniqueness of the event space within a given probabilistic model was put forward \cite{Robertson2005}. However the occurring ``wrong'' probabilities behaved sometimes neither in quantum, nor in classical way. The advances in quite different areas signal the need for a unified conceptual framework. An essential breakthrough in this direction was made by introducing topos approach to physical theories. In this essay possible applications of topos theory to Information retrieval are highlighted.

\section{Computational Complementarity} 

In this section primitive  {\em empirical statements} or {\em propositions} about automata \cite{df82} are introduced. Such  experimental statements form the basis of the formal investigation of the corresponding  logics. In particular, there exist automata for which validation of one empirical statement makes impossible the validation of another empirical statement and {\it vice versa}, as it was first pointed out by Moore \cite{moore}, which makes them similar to quantum systems.

Thereby, one decisive feature of the set-up is the {\em intrinsic} character of the measurement process: the automaton is treated as a black box with known description but unknown initial state. Automata experiments are conducted by applying an input sequence and observing the output sequence.

The conventional state identification problem \cite{moore} is to obtain information about an unknown initial state. One may think of it as choosing at random a single automaton from an automata ensemble which differ only by their initial state.  The task then is to find out which was the initial state of the chosen automaton.

Staying within the instrumentalist framework, we may ask if a classical object may demonstrate elements of quantum behavior. David Finkelstein provided a simple example of a classical automaton, which demonstrated quantum features on the level of the structure of the set of observable properties \cite{df82}. Like in quantum system, every measurement on its state affected the state itself, so the logic of accessible properties of Finkelstein's automaton with the graph

\medskip

\[
\unitlength 1.50mm
\begin{picture}(24,20)
\put(5,5){\line(0,1){15}}
\put(20,5){\line(0,1){15}}
\put(5,5){\line(1,0){15}}
\put(5,20){\line(1,0){15}}
\put(5,5){\circle*{2}}
\put(20,5){\circle*{2}}
\put(20,20){\circle*{2}}
\put(5,20){\circle*{2}}
\put(0,20){1}
\put(22,20){2}
\put(22,0){3}
\put(0,0){4}
\end{picture}
\]
was similar to that of a quantum particle with spin $1/2$.
\[
\unitlength 1.00mm
\linethickness{0.4pt}
\begin{picture}(113.28,55.00)
\put(62.33,10.00){\circle*{1.89}}
\put(62.33,50.00){\circle*{1.89}}
\put(52.33,30.00){\circle*{1.89}}
\put(32.66,30.00){\circle*{1.89}}
\put(92.66,30.00){\circle*{1.89}}
\put(72.66,30.00){\circle*{1.89}}
\multiput(62.33,50.00)(0.18,-0.12){167}{\line(1,0){0.18}}
\multiput(92.33,30.00)(-0.18,-0.12){167}{\line(-1,0){0.18}}
\multiput(62.33,10.00)(0.12,0.23){84}{\line(0,1){0.23}}
\multiput(72.33,29.67)(-0.12,0.24){84}{\line(0,1){0.24}}
\multiput(62.33,50.00)(-0.12,-0.23){84}{\line(0,-1){0.23}}
\multiput(52.66,29.67)(0.12,-0.24){81}{\line(0,-1){0.24}}
\multiput(62.33,10.00)(-0.18,0.12){164}{\line(-1,0){0.18}}
\multiput(32.66,29.67)(0.17,0.12){170}{\line(1,0){0.17}}
\put(62.33,5.00){\makebox(0,0)[cc]{$\emptyset$}}
\put(62.33,55.00){\makebox(0,0)[cc]{$\{1,2,3,4\}$}}
\put(18.00,30.00){\makebox(0,0)[lc]{$\{2,3,4\}$}}
\put(38,30.00){\makebox(0,0)[lc]{$\{1,3,4\}$}}
\put(73.66,30.00){\makebox(0,0)[lc]{$\{1,2,4\}$}}
\put(93.66,30.00){\makebox(0,0)[lc]{$\{1,2,3\}$}}
\end{picture}
\]
In the above example, it is not necessary to input sequences containing more than one symbol, since the non-final states are not distinguished by the output function. 

Since then a consistent theory of finite automata simulating quantum systems with a finite number of degrees of freedom was developed \cite{Grib90, Karl95}. A more general framework was developed on both technical and operationalistic level by Aerts with co-authors, including, in particular, study of cognition from a quantum perspective \cite{Aerts2009}.

\medskip

Anyway, on this step the emerging virtual structures was on the level of the logic of accessible properties, or, in other words, on the level of configuration space rather than fully fledged space-time.

\section{Classical spatiotemporal structures in Information Retrieval}

The basic idea of this approach, put forward in \cite{GBM} and developed in \cite{daniel} is the following. A smooth continuous manifold $\capb$ is considered, called {\em Information Retrieval space} (IR space).  Note that $\capb$ is neither a document space, nor a query space; instead, it has a more fundamental and unstructured nature. It may be thought of as the set of all transmitted bits. The elements of $\capb$ are all the same; they have no structure. This is a complete analogy to the points of spacetime, or the configuration space in physical theories.

To be more specific,$\capb$ is an analog to physical spacetime. Both documents and queries re placed into $\capb$, providing them with both temporal and spatial dimensions. As a consequence, the idea that a document may change in time is automatically incorporated in the theory. The second consequence is that the static documents, and those generated on-the-fly, are described as entities of exactly the same nature, differing only in `shape' in our IR spacetime $\capb$. Furthermore, since the points of $\capb$ are uniform, they can be treated as events. 

The IR process itself is represented by trajectories in $\capb$ which show the behavior of users. When we are speaking about a huge community of users, we no longer treat their behavior as intelligent: this gives us the right to shift the focus of our research from the task of finding a good way to satisfy users' requests to the task of analyzing typical user behavior. From this perspective, a typical user of a search environment is not more intelligent than an elementary particle or a molecule, and we may apply the good old principle of least action, which stems from the work of Fermat and Euler. This principle proved its efficiency by providing simple and strongly predictive models. 

The standard IR paradigm treats the IR process as a search. That is, the initial condition is posing a query, then, according to this or that formula, the indexed documents are ranked. Subsequently, the results are delivered to the user according to the ranking. But, typically, the user never makes a single query and the process is usually progressive. After parsing the results and considering their relevance, the user poses further queries, repeating the process iteratively based on relevance feedback.

Building IR space is getting above these particularities. The notion of relevance feedback is replaced by the least action: 
\begin{center}
\includegraphics[width=0.5\textwidth]{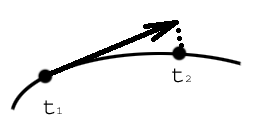}

Figure 1. A point on IR surface together with users' intention
vector.
\end{center}
The sequence of user's actions is interpreted as geodesic motion. The dynamics replace the notion of relevance, and the displacement of a user from point $t_1$ to point $t_2$ is what replaces relevance feedback making it, in a sense, a \emph{relevance feedforward}. 

Let us briefly overview how IR space is built on the basis of query log analysis.  Let us first produce the `flesh' of IR space. Its elementary constituent, a point, is a click: a query/HTTP request together with a body response (HTML page accessed by a result link). In order to specify a particular geodesic, we must specify the initial conditions which are the initial position $x(0)$ and the initial `intention' $\dot{x}(0)$. A typical user clickstream will be represented as a line, whereby the points of the line $x(t)$ are associated with the state of knowledge the user has gained from interpreting the retrieved information until that point.

  A clickstream is a progressive,  `continuous' sequence of user queries and responses which have a definite start and end.  The result of Step 1 is a collection of clickstreams, an ordered sequences of points:
\begin{center}
\includegraphics[width=0.4\textwidth]{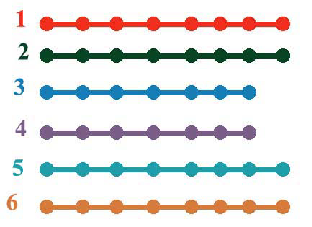}

Figure 2. Points on clickstreams are ordered by timestamps of
clicks.
\end{center}
At next step, the discrete pre-space $\capb$ is created. In the given  collection of clickstreams, their points are ordered and we know the distances between them. We assume that we use certain distance between points of the threads (any particular relevance formula can be applied). This way the clickstreams acquire metric:
\begin{center}
\includegraphics[width=0.5\textwidth]{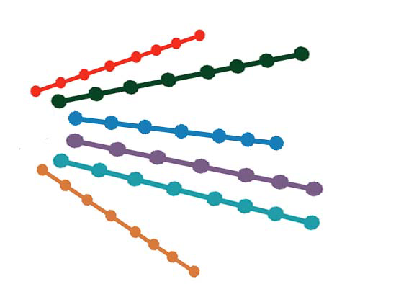}

Figure 3. Clickstreams acquire metrics.
\end{center}
The next step turns the structure from spatial to spatiotemporal by creating layers (analogs of spacelike hypersurfaces). Start with points with label 0 (this will be a starting layer), and, using the same distance function,  place them as points on a metric space. Then we pass to sequential label 1, and form the same discrete metric space, and so on. As a result, we have a sequence of layers labeled $0,1,\ldots$, forming altogether a discrete metric space:
\begin{center}
\includegraphics[width=0.5\textwidth]{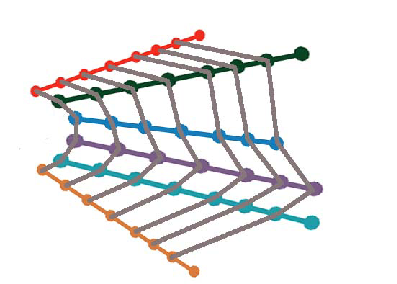}

Figure 4. Creating transversal layers.
\end{center}
The next step completes the skeleton with the `bones' linking nearest neighbors, now irrespective of the thread, to which they belong
\begin{center}
\includegraphics[width=0.5\textwidth]{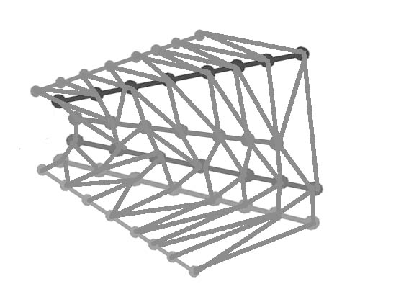}

Figure 5. Forming the discrete skeleton.
\end{center}
In the formalism proposed in \cite{daniel} the dimension $n+1$ of the future IR space is set {\em ad hoc} (since it has a spatiotemporal structure, one dimension is reserved for the temporal parameter and $n$ for `spatial'.)  Once $n$ is chosen, each layer is projected on an $n$-dimensional space, a foliation is formed, labeled  $1,2,\ldots$, together with threads, which are retained fromthe first step. Finally, the resulting space is treated as IR space.
\begin{center}
\includegraphics[width=0.7\textwidth]{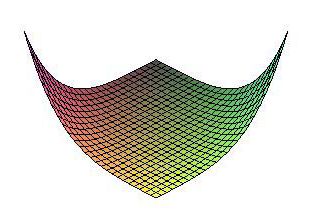}

Figure 6. IR space is built.
\end{center}

\section{A step towards quantum: adaptive artificial neural networks}

Artificial neural networks (ANNs) have gained increasing interest, finding applications in various domains such as business planning, medicine, engineering, geology, and physics. ANNs have shown remarkable efficiency in solving tasks related to forecasting, planning, control and classification.  ANNs are a natural and intuitive field to explore, as they draw inspiration from the structure of real biological systems. One of the basic tasks of neural networks is to function as perceptrons, that is, to recognize signals for which we have  no structural theory---for instance, to recognize visual patterns. 

We shall deal with multilayered feedforward NNs, such as 
\[
\unitlength0.7mm
\begin{picture}(100,40)
\multiput(0,0)(0,10){4}{\circle{2}} 
\multiput(20,10)(0,10){3}{\circle{2}} 
\multiput(40,0)(0,10){4}{\circle{2}} 
\multiput(0.75,10)(0,10){3}{\vector(1,0){18.5}} 
\multiput(0.75,0.75)(0,10){2}{\vector(2,1){18.5}} 
\put(0.75,10.75){\vector(1,1){18.5}} 
\multiput(20.75,9.25)(0,10){3}{\vector(2,-1){18.5}} 
\put(20.75,19.35){\vector(1,-1){18.5}} 
\multiput(20.75,10.75)(0,10){2}{\vector(2,1){18.5}} 
\put(20.75,30){\vector(1,0){18.5}} 
\end{picture}
\]
that is, their nodes can be arranged in layers so that (i) no nodes in a given layer communicate, and (ii) the signals propagate only consecutively, via layers. 

\paragraph{Training and performance.} Initially, one starts with a set of patterns for which the classification is known.  Usually by means of heuristic methods, the topological structure of the network is chosen and then trained via input of known patterns and subsequent adjustment of network parameters (transition functions). Output signals are correlated with patterns from different classes, to be well-separated with respect to certain criterion.  The most popular method to adjust transition functions is error backpropagation. Signal propagation in the linear approximation can be viewed as a matrix multiplication, which reduces to a number of arithmetic operations. The more links there are between neurons, the more computational resources are consumed by the process of pattern recognition. In order for a neural network to be faster, we should seek sparser configurations. Therefore, the criterion for `good matching' should also take performance into consideration.

\paragraph{From ANNs to AANNs.} The core idea of the suggested approach is to convert the primary component of artificial neural networks -- directed acyclic graphs (DAGs) -- into a more general structure, which naturally complies with quantum parallelism. The latter inevitably means that this should be a linear structure in order to support superpositions. In \cite{CQG-2002} the idea to replace DAG by appropriate sets of linear matrices (associated further with appropriate quantum variables) was suggested. The training paradigm remains the same as it was in conventional ANNs -- standard performance criteria are used, such as Root Mean Square Error (RMSE) between targets and outputs of the neural network, the correlation (called R-value) between the outputs and targets an so on, the are only reclassified in terms of unitary matrices and matrix operations. The main peculiarity is that the result of training optimization is then a matrix, or a set of matrices, rather than a directed acyclic graph, as is seen in a standard, `literal' approach. This way \textit{Adaptive Artificial Neural Networks (AANNs)} are introduced which, strictly speaking, are not networks. 

The topology of a feedforward artificial neural network, $\gnn$, is described by the \textit{template matrix}  of the appropriate directed acyclic graph (DAG), which is formed as follows:
\begin{equation}\label{etempl}
\ajm_{jk} \;=\; \left\lbrace
\begin{array}{l}
\ast,\:\mbox{if}\: j\to k \:\:\mbox{in}\:\: \gnn \cr
0,\:\mbox{otherwise}
\end{array}
\right.
\end{equation}
where $\ast$ stands for a wildcard -- any number, and the set of such numerical matrices form an algebra \cite{rota} as it is closed under multiplication. The main property of $\ajm$ is that the synaptic weights `follow' it -- namely, if $\ajm_{jk}=0$, then $w_{jk}=0$. This differs from the standard description in terms of adjacency matrices, and the difference is that the matrix is not numeric. The simplest example is provided by a two-vertex DAG with one arrow
\unitlength2mm 
\begin{equation}\label{ets}
\gnn=
\begin{picture}(30,1)(0,4.3)
\put(5,5){\circle{2}}
\put(6,5){\vector(1,0){17.5}}
\put(25,5){\circle{2}}
\end{picture}
\end{equation}
whose template matrix is
\begin{equation}\label{ema}
\ajm=\left(
\begin{array}{cc}
\ast & \ast \\
0 & \ast
\end{array} 
\right) 
\end{equation}
In general, taking various matrices by substituting the asterisk in template matrix \eqref{etempl} by numbers we may form various products. The resulting set of matrices is called the \textit{Rota algebra}. It can be verified that this set is closed under sums and matrix products, and thus qualifies as a closed algebra. This description is explicated under greater detail in \cite{CQG-2002}.

\paragraph{Rota topology.} The matrix formalism deals with objects having no specific spatiotemporal features. We call them continuously-evolving, superposed topologies, but these topologies are yet `spaceless' (while not `timeless,' as we are speaking of their evolution in time, with respect to the proper time of the experimenter). The possibility to transfer them into spacetime manifold fluctuations was studied in quantum gravity \cite{CQG-2002}. 

\medskip
\noindent The key point of the spatialization procedure \cite{CQG-2002} is the following: consider the full matrix algebra $\aaa$ as a linear space. From this perspective, the Rota algebras are just linear subspaces of $\aaa$. Having a procedure which -- starting from a subset of $\aaa$ -- creates a topological space, providing the capability to discuss superposed configurations of differing neural networks. In this section we present a procedure which -- starting from a given subspace of $\aaa$ -- produces a set, and endows it with a topology that can be associated with certain acyclic directed graphs.

In brief, the spatialization procedure makes it possible to transform matrix algebras of the form like \eqref{ema} to topological spaces like \eqref{ets}, so that they are endowed with the Rota topology. The emerging in this way classical spaces are the result of a quantum measurement, described by the Projection Postulate.  

\section{``Wrong'' probabilities: within and beyond the quantum}\label{swrongprob}

What do I mean by ``wrong'' probabilities? This is the numerical discrepancy between the predictions of a model based on Kolmogorov paradigm, that is, on the existence of a predefined sample space and the frequencies measured within an experiment. 

\paragraph{An example: Bell inequalities.} Consider three random variables $a,b,c$ each taking values $\pm 1$. Then whatever be their distributions, the following inequality holds:
\begin{equation}\label{ebell}
P(a=b)+P(b=c)+P(a=c)\ge 1
\end{equation}
The proof is straightforward: since all $a,b,c$ take only two values, at least one of the above equalities hold, therefore at least one of the (always nonnegative) summands in \eqref{ebell} equals to $1$. However, if $a,b,c$ are the measured projections of the spin of a quantum particle, it may happen that each summand in \eqref{ebell} equals to $\frac{1}{4}$. I do not dwell on this issue in detail, it gave rise to discussions and research lasting few recent decades, see, say, \cite{sep-bell-theorem} for a review. The only thing to conclude for this essay is that there is a realistic situation, in which the standard Kolmogorovian probabilistic model is not adequate. 

\paragraph{Generalities.} In order to test this or that model, Accardi's statistical invariants \cite{AccardiFedullo82} are employed, they allow to test the applicability of Kolmogorovian model. Given:

\begin{itemize}
\item a family of discrete maximal observables $\{A_\alpha : \alpha=1,\ldots T\}$ ($T$ being finite), each observable $A_\alpha$ takes the finite number of values $a^{(\alpha)}_{j\alpha}$ labeled by $j_\alpha=1,\ldots,n$
\item the experimentally measurable conditional probabilities $p_{j_\alpha,j_\beta}(\beta\mid\alpha)$
\begin{equation}\label{eacc32}
p_{j_\alpha,j_\beta}(\beta\mid\alpha)
\;=\;
P\left(A_\beta = a^{(\beta)}_{j \beta}\right\rvert
\left.A_\alpha = a^{(\alpha)}_{j \alpha}\right)
\end{equation}
\end{itemize}
The problem is: does there exist a probability space $(\samps ;\eventa ; P)$ and $T$ measurable partitions $A_{j}^{(\alpha)}$ of cardinality $n$ (the number of distinct values of each observable is assumed to be the same)
\begin{equation*}\label{eaccaa}
A_{j}^{(\alpha)}, \alpha=1,\ldots T,\,j=1,\ldots n
\end{equation*}
such that for any $\alpha,\beta=1,\ldots T$ one has
\begin{equation}\label{eacckolm}
P\left(A^{(\beta)}=a_{j}^{(\beta)} \mid A^{(\alpha)}=a_{i}^{(\alpha)}\right)
\;=\;
\frac{P\left(A_{j}^{(\beta)}\cup A_{i}^{(\alpha)}\right)}{P\left(A_{j}^{(\beta)}\right)}
\end{equation}
In general, this is a finitely decidable linear programming problem \cite{AccardiFedullo82}. As an example, consider the special case of three observables $A,B,C$, each taking only two values $a_1,a_2$ for $A$, $b_1,b_2$ for $B$ and $c_1,c_2$ for the observable $C$. The transition probability matrices for each pair of observables, being bistochastic, each has only one numeric parameter, denote the appropriate matrices as
\begin{equation}\label{ebistoch}
\begin{array}{l@{\;=\;}c@{\;=\;}c}
P(A\mid B)&P&\left(%
\begin{array}{cc}
p & 1-p \\
1-p & p \\
\end{array}%
\right) \\
P(B\mid C)&Q&\left(%
\begin{array}{cc}
q & 1-q \\
1-q & q \\
\end{array}%
\right) \\
P(C\mid A)&R&\left(%
\begin{array}{cc}
r & 1-r \\
1-r & r \\
\end{array}%
\right) \\
\end{array}
\end{equation}
then these transition probabilities can be described by a Kolmogorovian model (that is, they are produced by a single sample space) if and only if
\begin{equation}\label{eaccardi}
\lvert p+q-1\rvert
\leq r \leq
1-\lvert p-q \rvert
\end{equation}
That means, when we use probabilities derived from relative frequencies, which, in the meantime, can be calculated from other given probabilities, we can directly test the inequality \eqref{eaccardi}. If it fails, that signals that we deal with ``wrong'', non-classical probabilities. They cannot be described in terms of a single event space and thus require a notion of a `varying event space'.

\paragraph{Non-classical probabilities in Information Retrieval. } In the realm of Information Retrieval, the probabilities are usually obtained from various sources, and, as it happens, they may not admit a single event space. ``Not admit'' means that it is not possible to assign a single elementary event space and a probability measure satisfying Kolmogorovian axioms. 

In practice, relative frequencies rather than probabilities are measured. Then, the relative frequencies are thought of as probabilities and, looking at the values of those probabilities, decisions are made. Are these decisions good or bad? What can we change in order to enhance the performance of search environment? The point is that for some probabilities, there are different ways to be obtained: the value can be either

(i) obtained \emph{directly} by measuring appropriate relative frequency, or 

(ii) \emph{derived} from the values of other observed relative frequencies. 

\medskip

The latter needs an underlying probabilistic model. Supposed it is Kolmogorovian, so the existence of a (single) sample space $\samps$ is required. The events are subsets of $\samps$, while the points of the sample space are elementary and independent. 

The queries are done multiply. We are going to make evaluations based on ratios. The standard picture is:

\mbox{
\unitlength.8mm
\begin{picture}(180,110)
\put(0,0){\framebox(140,100)[tr]{}}
\put(30,90){$\samps$}
\put(10,10){\framebox(70,70){\mbox{Relevant}}}
\put(62,20){\framebox(70,70){\mbox{Retrieved}}}
\end{picture}
}

The key problem is that the sample space $\samps$ is not well defined. This suggests the notion of varying, context-dependent sample space. 

\paragraph{Melucci metaphor.} Melucci metaphor is a unified view to represent simplified IR environment with no reference to particular underlying logic and rendering IR into experimental, naturalist realm. According to it \cite{Melucci2012}, the IR procedure is represented by a two-slit experiment, widely known in physics. The IR system is thought of as a laboratory with the source, which supplies documents according to the input query. What is described, looks intermediate between a pure Gedankenexperiment and a fully fledged measurement.

The documents within Melucci metaphor are treated as particles, they may be of classical or quantum nature, or, perhaps, of some other kind. We do specify the mechanism of producing this flux of documents-particles. What is essential, is that the number of ejected documents is supposed to be potentially infinite, but we analyze only first $N$ documents. Entering a particular query means preparing the source in a particular state $\qss$. From this experiment we get the value of $P(\ppx\vert\rlv)$, where $P(\ppx)$ stands for the probability of the occurrence of a certain term $\ppx$:
\unitlength.8mm
\[
\begin{picture}(180,60)
\put(0,30){\oval(40,40)[r]}
\put(21,33){\vector(2,1){15}}
\put(22,32){\vector(4,1){14.5}}
\put(22,30){\vector(1,0){14}}
\put(22,28){\vector(4,-1){14.5}}
\put(21,27){\vector(2,-1){15}}
\put(50,3){\line(0,1){8}}
\put(50,19){\line(0,1){16}}
\put(50,43){\line(0,1){8}}
\multiput(49,11)(0,8){2}{\line(1,0){2}}
\put(52,21){$\nrlv$}
\multiput(49,35)(0,8){2}{\line(1,0){2}}
\put(52,44){$\rlv$}
\multiput(48.8,10)(0.1,0){4}{\line(0,1){10}}
\multiput(57,36)(0,3){3}{\vector(1,0){14}}
\put(77,24){\framebox(27,27){Check $\ppx$}}
\end{picture}
\]
and from this experiment we get the value of $P(\ppx\vert\nrlv)$:
\[
\begin{picture}(180,55)
\put(0,30){\oval(40,40)[r]}
\put(21,33){\vector(2,1){15}}
\put(22,32){\vector(4,1){14.5}}
\put(22,30){\vector(1,0){14}}
\put(22,28){\vector(4,-1){14.5}}
\put(21,27){\vector(2,-1){15}}
\put(50,3){\line(0,1){8}}
\put(50,19){\line(0,1){16}}
\put(50,43){\line(0,1){8}}
\multiput(49,11)(0,8){2}{\line(1,0){2}}
\put(52,21){$\nrlv$}
\multiput(49,35)(0,8){2}{\line(1,0){2}}
\put(52,44){$\rlv$}
\multiput(48.8,34)(0.1,0){4}{\line(0,1){10}}
\multiput(57,12)(0,3){3}{\vector(1,0){14}}
\put(77,3){\framebox(27,27){Check $\ppx$}}
\end{picture}
\]
When we are in the classical realm, there is no need to calculate $P(\ppx)$ due to our Boolean belief revision (that is, the law of total probability):
\begin{equation}\label{eltp}
P(\ppx)=P(\ppx\vert\rlv)\,P(\rlv)+P(\ppx\vert\nrlv)\,P(\nrlv)
\end{equation}
But just for fun we may attempt to measure $P(\ppx)$ directly, removing the relevance check:
\[
\begin{picture}(180,55)
\put(0,30){\oval(40,40)[r]}
\put(21,33){\vector(2,1){15}}
\put(22,32){\vector(4,1){14.5}}
\put(22,30){\vector(1,0){14}}
\put(22,28){\vector(4,-1){14.5}}
\put(21,27){\vector(2,-1){15}}
\put(50,3){\line(0,1){8}}
\put(50,19){\line(0,1){16}}
\put(50,43){\line(0,1){8}}
\multiput(49,11)(0,8){2}{\line(1,0){2}}
\put(52,21){$\nrlv$}
\multiput(49,35)(0,8){2}{\line(1,0){2}}
\put(52,44){$\rlv$}
\multiput(57,36)(0,3){3}{\vector(1,0){14}}
\multiput(57,12)(0,3){3}{\vector(1,0){14}}
\put(77,11){\framebox(33,34){Check $\ppx$}}
\end{picture}
\]
and, surprisingly, discover that the result may drastically differ from \eqref{eltp}. Let us pass to exact numerical results. In order to evaluate the discrepancy, Accardi statistical invariant is used:
\begin{equation}\label{eaccdef}
\asi=
\frac{P(\ppx)-P(\ppx\vert\nrlv)}{P(\ppx\vert\rlv)-P(\ppx\vert\nrlv)}
\end{equation}
When the IR environment is classical, the law of total probability \eqref{eltp} holds, therefore
\[
\asi=
\frac{P(\ppx\vert\rlv)P(\rlv)+P(\ppx\vert\nrlv)P(\nrlv)-P(\ppx\vert\nrlv)}{P(\ppx\vert\rlv)-P(\ppx\vert\nrlv)}=
\]
\[
=
\frac{P(\ppx\vert\rlv)P(\rlv)-P(\ppx\vert\nrlv)P(\rlv)}{P(\ppx\vert\rlv)-P(\ppx\vert\nrlv)}=P(\rlv)
\]
that is why
\[
0\leq \asi \leq 1
\]
in classical realm. In quantum setting this inequality may be violated.

At first sight, a violation of Accardi inequality \eqref{eaccardi} seems to provide wrong results since a ``wrong'' model of the world cannot provide nothing but ``wrong'' results. However, the experiments were performed showing that the violation can
enhance the search effectiveness~\cite{Melucci10}. It was observed that the terms, whose {\em measured} probabilities violates~\eqref{eaccardi}, are those
that increase average precision more frequently and significantly than those do not.

So, it is conjectured that a single-event-space model may be insufficient. As already was mentioned, a theoretical support for the notion of `varying event space' is needed. 

\section{Non-classical models of Information retrieval}\label{stopoi}

The notion similar to that of configuration space in Classical Mechanics or event space in Probability Theory is compelling for a theory to be realistic. The rough reason is that the events need a room to occur. Although, from both theoretical and pragmatic points of view, one can move from realism to instrumentalism. Instead of telling that a system possesses this or that property, we say that the measurement yields this or that result. This result we may (or may not) be interpreted as property. But nothing but results are available. When we qualify a system as classical and quantum, this is not about its `real' nature. They are about the observable properties. In particular, probabilities need not inevitably be interpreted as a reflexion of statistics, of relative frequencies. Rather, they can be viewed as propensities, therefore they need not be a number between 0 and 1, they may form a more complicated structure. This idea is developed within the topos approach to Quantum Mechanics. 

\medskip 

Nowadays, a topos approach to a theory of quantum gravity is well-developed \cite{doering-isham-rev}, the idea on how topos theory can be used in general to describe theories of physics includes, in particular, a theory of quantum gravity. Physical theories are formulated in a topos other than $\Sets$, this topos may depend on both the theory-type and the system. The notion of {\em theory-type} is crucial, this will be further expanded from the realm of ``good old'' physics to that of Information Retrieval. 

If a theory-type, such as classical physics, or quantum physics, or something else is applicable to a certain class of systems, then, for each system in this class, there is a topos in which the theory is to be formulated. For some theory-types the topos is system-independent: for instance, classical physics always uses the topos of sets. For other theory-types, the topos may vary from one system to another; this is, for instance, the case for quantum mechanics.

The main particular goal is finding, in a topos, a representation of a certain formal language, that is, associated with the system in question.

\paragraph{Non-classical models of Information retrieval.} The experimental demonstrate violations of seemingly necessary theoretical invariants. The suggestion is to treat this as a potential way to improve the IR effectiveness. The idea to use quantum model was put forward by van Rijsbergen \cite{vanRijsbergen04}. In the meantime, there is no evidence that other inequalities, perhaps even non-quantum, can be formulated that their violation can reveal, from the one hand, the inconsistency between experimental observations and a particularly chosen underlying model, but, on the other hand, the directions toward a further development of models for IR. Still, the question remained why the abstract vector spaces would be a better framework than other mathematical theories. A possible answer was provided in~\cite{vanRijsbergen04}: Hilbert spaces encompass different models for information retrieval, such as the probabilistic model and the Vector Space Model. This proved the use of non-classical logic to be effective. But non-classical does not necessarily means quantum, and a broader conceptual background is needed to build a uniform theory. The program of building an operationalistic framework for empirically verified theories (we may call them physical theories, but they should be understood in a broader sense) was successfully carried out in the last decade, see \cite{doering-isham-rev,Flori2012} for reviews.

Topos theory is a vast, broadly developed area of mathematics, there is no need to define it in detail in this essay. A very good introduction to the subject is \cite{GoldblattBook}, the reviews of applications of topos theory to physics can be found in a general context \cite{doering-isham-rev} and specifically for Quantum Gravity in \cite{Flori2012}. 

A loose definition of a topos is a category with special properties, which make a topos ``look like'' the category of sets, $\Sets$, in the sense that any mathematical operation that can be done in set theory has an analogue in a general topos. What is essential is that the conceptual needs for both theoretical physics and Information Retrieval are in accordance: 
\begin{itemize}
\item Spatialisation (in whatever sense) is needed in order to make theory realistic and experimentally verifiable
\item In order to be in a position to {\em dasein}, a kind of classical snapshot of a system is to be taken. 
\item To create the arena for events, a Gel'fand-like spatialisation is required
\item To obtain a reasonable Gel'fand spectrum, a commutative subalgebra, associated with a particular context, should be specified.
\end{itemize}
This scheme is common for building a probabilistic-like model in any operationalist environment. To conclude, topos paradigm provides a formalism, which, given a physical system (or, more generally, an operationalistic environment) associates with it a bunch of virtual context-dependent configuration (or sample) spaces. 

\section*{Concluding remarks}

We see that the experimental results over large data collections demonstrate features of quantum behavior. In order to mimic quantum indeterminacy, the access to complete knowledge about the system was artificially restricted. But this phenomenon is generic for Information Retrieval! The point is that search engines store some limited data about the documents rather than the documents themselves. This is a natural restriction for the access to the documents to be complete, which, in turn, could be the reason for the observed non-classicality. 

The task for IR is to range search results. The standard probability-based approach may be too straightforward and not effective sometimes. A more subtle evaluations like propensities are invoked. These propensities must be somehow quantified. The topos approach is flexible enough to provide various scales for propensities. Within it, ``probabilities'' may not lie with the $[0,1]$ interval, furthermore, it admits models where ``probabilities'' are not totally ordered. The existence of non-comparable properties is natural for IR. But these are practical issues, which I do not inquire here. 

To conclude, I would like to formulate the main message of this essay. There are different physical theories. To evaluate them, one needs to know what each of these theories has to do with {\em the} reality. My claim is that nowadays we witness the creation of realities, these realities become to greater and greater extent real in comparison with {\em the} reality. As a consequence, we are now in a position to say that certain theory describes {\em a} reality. What happens with a physical theory if it turns out that its predictions are disproved by experiment? The theory is to be rejected. The reason to reject it is that it does not describe the reality. In this essay I show that nowadays we witness how a new harbor for the abandoned theories is being built. The factory producing these realities increases its output, and, therefore, abandoned physical theories, which were proved to be falsified by experiment, may find the new life. 

\paragraph{Acknowledgments.} The author appreciates Cris Calude, Karl Svozil and Jozef Tkadlec for stimulating discussions on quantum contextuality during my stay in Technical University of Vienna, supported by the TU Wien Ausseninstitut and the Institute of Theoretical Physics of the Vienna University of Technology. I am also grateful to Petros Wallden and other participants of the Working Group Meeting `Foundations of Quantum Mechanics and Relativistic Spacetime' for COST Action MP1006, 25-26 September 2012, University of Athens, Greece.

\end{document}